\documentclass[twoside]{article}
\usepackage{graphicx}
\usepackage{fancyhdr}
\usepackage{multicol}
\usepackage{styl-ap}

\begin{document}
\title{ARGO-YBJ: Status and Highlights}
\author{Giuseppe Di Sciascio\work{1} on behalf of the ARGO-YBJ Collaboration}
\workplace{INFN - Sezione di Roma Tor Vergata, Viale della Ricerca Scientifica 1, I-00133 Roma}
\mainauthor{disciascio@roma2.infn.it}
\maketitle

\begin{abstract}%

The ARGO-YBJ experiment is in stable data taking since November 2007 at the YangBaJing Cosmic Ray Laboratory (Tibet, P.R. China, 4300 m a.s.l., 606 g/cm$^2$).
ARGO-YBJ is facing open problems in Cosmic Ray (CR) physics in different ways. The search for CR sources is carried out by the observation of TeV gamma-ray sources both galactic and extra-galactic. 
The CR spectrum, composition and anisotropy are measured in a wide energy range (TeV - PeV) thus overlapping for the first time direct measurements.
In this paper we summarize the current status of the experiment and describe some of the scientific highlights since 2007.
\end{abstract}


\begin{multicols}{2}
\section{Introduction}

Exploiting the full coverage approach at high altitude, the ARGO-YBJ experiment is an air shower array able to detect the cosmic radiation with an energy threshold of a few hundred GeV. 
The detector is in stable data taking since November 2007 with a duty cycle larger than 86\%. The trigger rate is 3.5 kHz. The detector characteristics and performance are described in detail in \cite{aielli06,bartoli11a}.
The main results obtained by ARGO-YBJ are described in \cite{dettorre11}. In this paper the status of the experiment is summarized and some highligths reviewed.

\section{Gamma-ray astronomy}

From November 2007 the ARGO-YBJ experiment collected about 4$\times$10$^{11}$ events in 1543 days of total effective observation time.
Five known VHE $\gamma$-ray sources have been detected with a statistical significance greater than 5 standard deviations (s.d.), i.e. Crab Nebula, Mrk421, Mrk501, MGRO J1908+06 and MGRO J2031+41. A number of flares from Mrk421 and Mrk501 were observed and studied in detail. Evidence of a TeV flaring activity from the Crab Nebula in coincidence with AGILE/Fermi observations is also reported.
Details on the analysis procedure (e.g., data selection, background 
evaluation, systematic errors) are discussed in \cite{aielli10a,bartoli11b}.

\subsection{Crab Nebula}

With all the data recorded in 3.5 years ARGO-YBJ observed a TeV signal from the Crab Nebula with a statistical significance of about 17 s.d., proving that the cumulative sensitivity of the detector reached 0.3 Crab units.

The observed flux is consistent with a steady emission, and the observed differential energy spectrum in the 0.3-30 TeV range can be described by
dN/dE = (3.0$\pm$0.30) $\times$ 10$^{-11}$ (E/TeV)$^{-2.57\pm 0.09}$ 
photons cm$^{-2}$ s$^{-1}$ TeV$^{-1}$, in good agreement with other observations. According to MC simulations, 84\% of the detected events comes from primary photons with energies greater than 300 GeV, while only 8\% comes from primaries above 10 TeV. We evaluate a systematic error on the flux less than 30$\%$ mainly due to the background estimaation and to the uncertainty on the absolute energy scale.

According to the AGILE and Fermi data 
4 major flaring episodes at energies E $>$100 MeV occurred during the ARGO-YBJ data acquisition \cite{tavani11,abdo11,striani11,ojha12}. 

{\bf Flare 1}: starting time MJD 54857 (Feb. 2009), duration $\Delta$t$\sim$16 days, maximum flux F$_{max}$ $\sim$5 times larger than the steady flux
\cite{abdo11}. During this flare no excess is present in our data, for any multiplicity threshold.

{\bf Flare 2}: starting time MJD 55457 (Sept. 2010), duration $\Delta$t $\sim$4 days, maximum flux F$_{max}$ $\sim$5 times larger than the steady flux \cite{tavani11,abdo11}. According to temporal data analysis the $\gamma$-ray emission is concentrated in 3 narrow peaks of $\sim$12 hours duration each \cite{balbo11,striani11}. Integrating the 3 transits we observe for N$_{pad}>$40 an excess of 3.1 s.d. over the expected steady flux (0.55 s.d.). 
If the excess were due to a flare, the $\gamma$-ray flux would be higher 
by a factor $\sim$5 with respect to the steady flux at energies around 1 TeV.
Integrating the data over 10 transits (from MJD 55456/57 to MJD 55465/66) the signal significance is 4.1 s.d. (pre-trial), while 1.0 s.d. is expected from the steady flux \cite{aielli10b}. 
No measurement from Cherenkov telescopes is available in coincidence with our observations and with the different spikes to confirm this excess. Sporadic measurements performed by MAGIC and VERITAS telescopes at different times from MJD 55456.45 to MJD 55459.49 show no evidence for a flux variability \cite{mariotti10,ong10}.

{\bf Flare 3}: starting time MJD 55660 (Apr. 2011) \cite{fermi3}, duration $\Delta$t $\sim$6 days, maximum flux F$_{max}$ $\sim$14 times larger than the steady flux \cite{striani11}.
Integrating the ARGO-YBJ data over the 6 days in which AGILE detected a
flux enhancement, i.e. from MJD 55662 to MJD 55668, we find evidence for an excess for events with N$_{pad}>$ 100 at a level of about 3.5 s.d. . No measurement from Cherenkov telescopes is available during these days,
due to the presence of the Moon during the Crab transits.

{\bf Flare 4}: starting time MJD 56111 (July 3rd, 2012) \cite{ojha12}, duration $\Delta$t $\sim$3 days. The daily-averaged emission doubled from (2.4$\pm$0.5)$\times$ 10$^{-6}$ ph cm$^{-2}$ sec$^{-1}$ on July 2nd to (5.5$\pm$0.7)$\times$10${-6}$ ph cm$^{-2}$ sec$^{-1}$ on July 3rd.
A preliminary analysis of the ARGO-YBJ data shows an excess of events with statistical significance of about 4 s.d. from a direction consistent with the Crab Nebula on July 3rd, corresponding to a flux $\approx$ 8 times higher than the average emission at a median energy of $\sim$1 TeV \cite{bartoli12a}. The expected steady flux corresponds to 0.33 s.d. . No significant exces is detected in the following days from July 4th to 6th. 
Once again, no measurement from Cherenkov telescopes is available during these days.

In conclusion, the ARGO-YBJ data show a marginal evidence of a TeV flux increase correlated to the MeV-GeV flaring activity, with a statistical significance not enough to draw a firm conclusion (the post-trial chance probability is of order 10$^{-3}$ for each flare). 
Nevertheless, the chance probability of observing 3 flares out of 4 in coincidence with the satellites is greatly depressed.
A detailed analysis of the Crab Nebula TeV emission is under way and a paper is in preparation.
\subsection{The blazar Mrk421}
ARGO-YBJ is monitoring Mrk421 above 0.3 TeV, studying the correlation of the TeV flux with X-ray data. We observed this source with a total significance of about 14 s.d., averaging over quiet and active periods. As it is well known, this AGN is characterized by a strong flaring activity both in X-rays and in TeV $\gamma$-rays. Many flares are observed simultaneously in both bands. The $\gamma$-ray flux observed by ARGO-YBJ has a clear correlation with the X-ray flux. No lag between X-ray and $\gamma$-ray photons longer than 1 day is found. The evolution of the spectral energy distribution (SED) is investigated by measuring spectral indices at four different flux levels. Spectral hardening is observed in both X-ray and $\gamma$-ray bands. The $\gamma$-ray flux increases quadratically with the simultaneously measured X-ray flux.
All these observational results strongly favor the Self-Synchro Compton (SSC) process as the underlying radiative mechanism.
The results of Mrk421 long-term monitoring are summarized in \cite{bartoli11c}.
\subsection{The blazar Mrk501}
The long-term observation of the Mrk501 TeV emission by ARGO-YBJ can be described by the following differential energy spectrum:
dN/dE = (1.92$\pm$0.44) $\times$ 10$^{-12}$ (E/2 TeV)$^{-2.59\pm0.27}$ 
photons cm$^{-2}$ s$^{-1}$ TeV$^{-1}$, corresponding to 0.312$\pm$0.076 Crab units above 1 TeV.

In October 2011 started the largest flare since 2005. During the brightest $\gamma$-ray flaring episodes from October 17 to November 22, 2011, an excess of the event rate over 6 s.d. has been detected by ARGO-YBJ, corresponding to an increase of the $\gamma$-ray flux above 1 TeV by a factor of 6.6$\pm$2.2 from its steady emission \cite{bartoli12b}. During the flare, the differential flux is dN/dE = (2.92$\pm$0.52) $\times$ 10$^{-12}$ (E/4 TeV)$^{-2.07\pm0.21}$ 
photons cm$^{-2}$ s$^{-1}$ TeV$^{-1}$, corresponding to 2.05$\pm$0.48 Crab units above 1 TeV.
Remarkably, $\gamma$-rays with energies above 8 TeV are detected with a statistical significance of about 4 s.d., which did not happen since
the 1997 flare. 
The average SED for the steady emission is well described by a simple one-zone SSC model. However, the detection of gamma-rays above 8 TeV during
the flare challenges this model due to the hardness of the spectra \cite{bartoli12b}. 
%
\begin{myfigure}
\centerline{\resizebox{70mm}{!} {\includegraphics{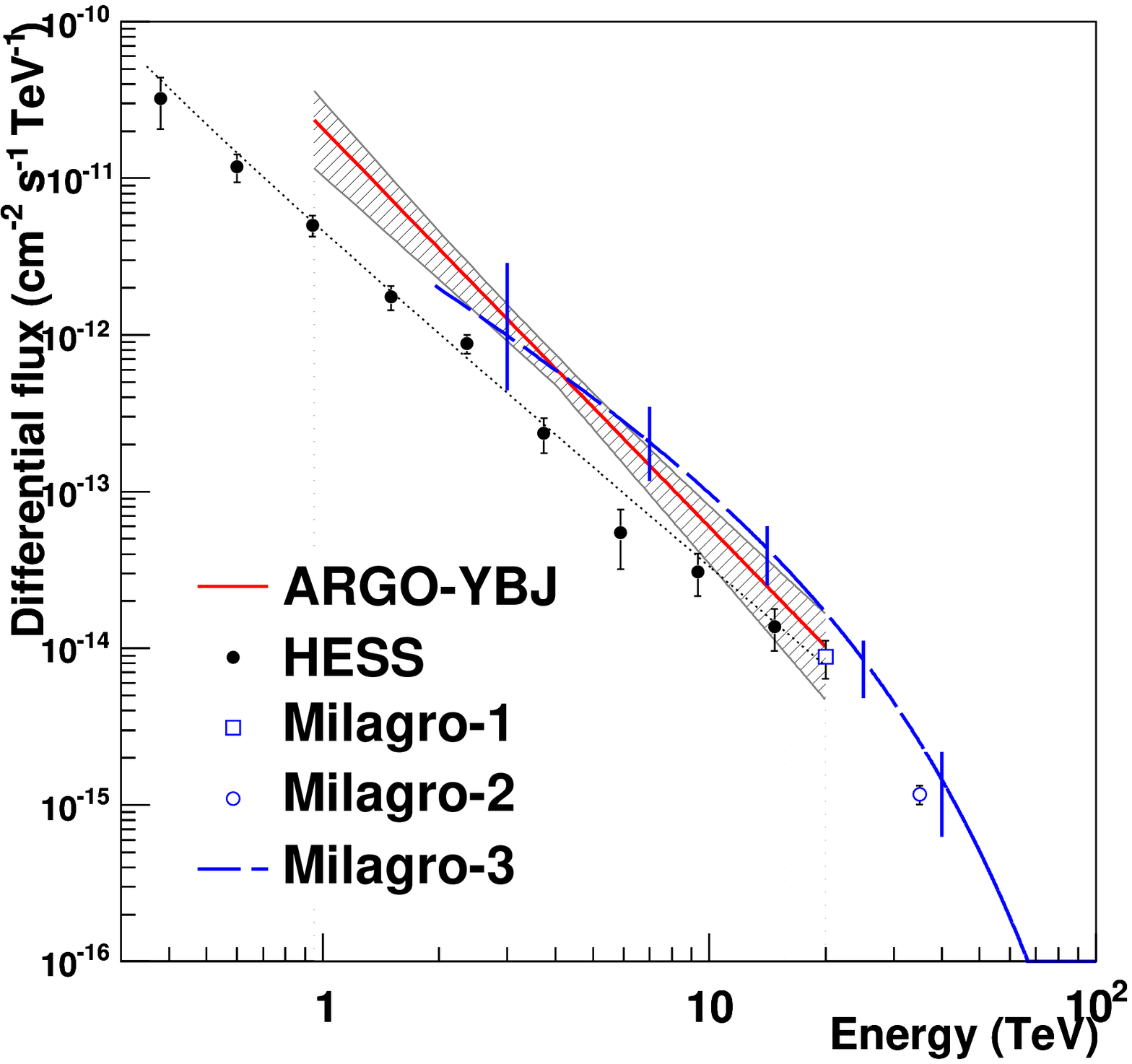}}}
\caption[h]{Gamma-ray flux from MGRO J1908+06 measured by ARGO-YBJ (red line) compared to other measurements. The dashed area represents the 1 s.d. error.
The plotted errors are purely statistical for all the detectors.
For details and references see \cite{bartoli12c}.}
\label{fig:mgro1908}
\end{myfigure}
%
%
\subsection{MGRO J1908+06}
The $\gamma$-ray source MGRO J1908+06 was discovered by Milagro at a median energy of $\sim$20 TeV and confirmed by HESS at energies above 300 GeV.
The Milagro and HESS energy spectra are in disagreement, being the Milagro result about a factor 3 higher at 10 TeV.

ARGO-YBJ observed a TeV emission from MGRO J1908+06 with a maximum significance of about 7.3 s.d. for N$_{pad} \geq$20 in 6867 hours on-source \cite{bartoli12c}. 
The intrinsic extension is determined to be $\sigma_{ext}$ = 
0.49$^{\circ}$$\pm0.22$, consistent with the HESS measurement ($\sigma_{ext}$ = 0.34$^{\circ}$$_{-0.03}^{+0.04}$). The best fit spectrum obtained is:
dN/dE = (6.1$\pm$1.4) $\times$ 10$^{-13}$ (E/4 TeV)$^{-2.54\pm0.36}$ 
photons cm$^{-2}$ s$^{-1}$ TeV$^{-1}$, in the energy range 1-20 TeV (see Fig. \ref{fig:mgro1908}).
The measured gamma-ray flux is consistent with the Milagro results, but is $\sim$2-3 times larger than the flux derived by HESS at energies of a few TeV. Given the reduced significance of the excess at high energies, we are not able to constrain the shape of the spectrum above 10 TeV and to definitively rule out a possibile high energy cutoff.

The continuity of the Milagro and ARGO-YBJ observations and the stable excess
rate observed by ARGO-YBJ along 4 years of data taking support the 
identification of MGRO J1908+06 as a stable extended source, likely the TeV nebula of PSR J1907+0602, with a flux at 1 TeV $\sim$67$\%$ that of the Crab Nebula. Assuming a distance of 3.2 kpc, the integrated luminosity above 1 TeV is $\sim$1.8 times  that of the Crab Nebula, making  MGRO J1908+06 one of the most luminous Galactic gamma-ray sources at TeV energies \cite{bartoli12c}.

\begin{myfigure}
\centerline{\resizebox{80mm}{!} {\includegraphics{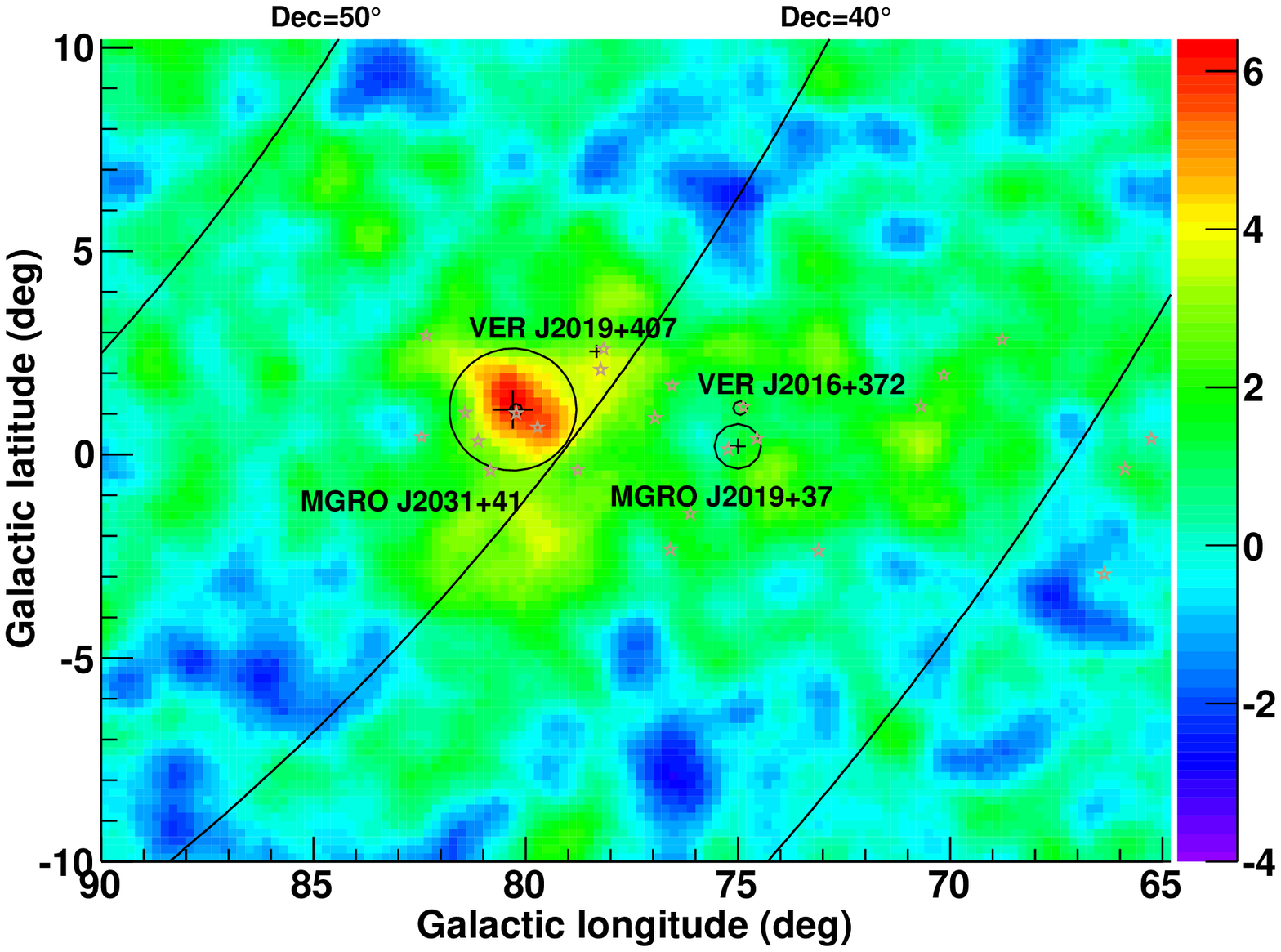}}}
\caption[h]{Significance map of the Cygnus region as observed by ARGO-YBJ for N$_{pad}>$20. The four known VHE $\gamma$-ray sources are reported.
  The errors on the MGRO source positions are marked with crosses, while the circles indicate their intrinsic sizes. The cross for VER J2019+407 indicates its extension. The source  VER J2016+372 is marked with a small circle without position errors. The small circle within the errors of MGROJ2031+41 indicates position and extension of the source TeV J2032+4130 as estimated by the MAGIC collaboration. The open stars mark the location of the 24 GeV sources in the second $Fermi$ LAT catalog. For details and references see \cite{bartoli12d}.}
\label{fig:cygnus}
\end{myfigure}
%
\subsection{MGRO J2031+41 and the Cygnus region}

The Cygnus region contains a large column density of interstellar gas and is rich of potential CR acceleration sites as Wolf-Rayet stars, OB associations and supernova remnants. Several VHE gamma-ray sources have been discovered within this region in the past decade,including two bright extended sources detected by the Milagro experiment. 

\begin{myfigure}
\centerline{\resizebox{80mm}{!} {\includegraphics{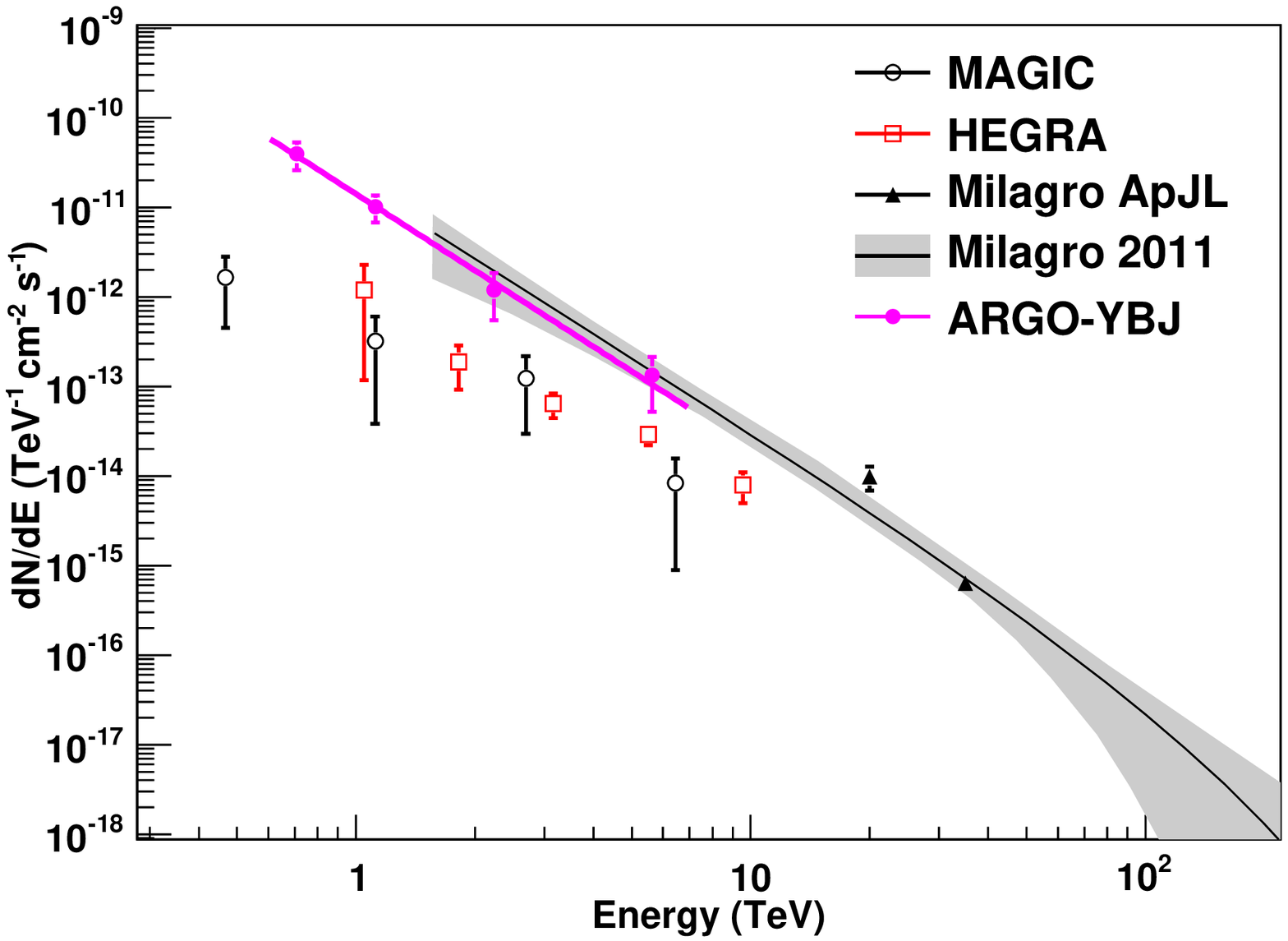}}}
\caption[h]{Energy spectrum of MGRO J2031+41/TeV J2032+4130 as
measured by ARGO-YBJ (magenta solid line). The spectral
measurements of HEGRA and MAGIC are also reported for comparison. The black solid line and shaded area indicate the differential energy spectrum and the 1 s.d. error region as recently determined by the Milagro experiment. The two triangles give the previous flux measurements by Milagro at 20 TeV and
35 TeV. For details and references see \cite{bartoli12d}. }
\label{fig:mgro2031}
\end{myfigure}
%

The gamma ray source MGRO J2031+41, detected by Milagro at a median energy of $\sim$ 20 TeV, is spatially consistent with the source TeV J2032+4130 discovered by the HEGRA collaboration and likely associated with the Fermi pulsar 1FGL J2032.2+4127.
The extension measured by Milagro, 3.0$^{\circ}\pm$0.9$^{\circ}$, is much larger than that initially estimated by HEGRA (about 0.1$^{\circ}$).  

The bright unidentified source MGRO J2019+37 is the most significant source in the Milagro data set apart from the Crab Nebula. This is an enigmatic source due to its high flux  not being confirmed by other VHE gamma-ray detectors.
Recently, a deep survey carried out by VERITAS with a sensitivity of $\sim$1\% Crab units showed a complex emitting region with different faint sources inside the MGRO J2019+37 extension. The estimated flux is much weaker than that determined by Milagro.

The Cygnus region has been studied by ARGO-YBJ with data collected in a total effective observation time of 1182 days \cite{bartoli12d}. The results of the data analysis are shown in Fig. \ref{fig:cygnus} and Fig. \ref{fig:mgro2031}.
A TeV emission from a position consistent with MGRO J2031+41/TeV J2032+4130 is found with a significance of 6.4 s.d. .  
Assuming the background spectral index -2.8, the intrinsic extension is determined to be $\sigma_{ext}=(0.2_{-0.2}^{+0.4})^{\circ}$, consistent with the estimation by the MAGIC and HEGRA experiments, i.e., (0.083$\pm$0.030)$^{\circ}$ and (0.103$\pm$0.025)$^{\circ}$, respectively.

The differential flux (Fig. \ref{fig:mgro2031}) is 
dN/dE = (1.40$\pm$0.34) $\times$ 10$^{-11}$ (E/TeV)$^{-2.83\pm0.37}$ 
photons cm$^{-2}$ s$^{-1}$ TeV$^{-1}$, in the energy range 0.6 - 7 TeV.
Assuming $\sigma_{ext}$ = 0.1 the integral flux is 31\% that of the Crab at energies above 1 TeV, which is higher than the flux of TeV J2032+4130 as determined by HEGRA (5$\%$) and MAGIC (3$\%$). Again, this measurement is in fair agreement with the Milagro result.

The reason for the large discrepancy between the fluxes measured by
Cherenkov telescopes and EAS arrays (ARGO-YBJ and Milagro) is still unclear. 
Possible contributions from diffuse $\gamma$-ray emission, nearby sources, and systematic uncertainties are not enough to explain the discrepancy \cite{bartoli12d}.

No evidence of a TeV emission above 3 s.d. is found at the location of MGRO J2019+37 and flux upper limits at 90\% c.l. are set. 
At energies above 5 TeV the ARGO-YBJ exposure is still insufficient to reach a firm conclusion while at lower energies the ARGO-YBJ upper limit is marginally consistent with the spectrum determined by Milagro. 
The observation by ARGO-YBJ is about five years later than that by Milagro. A flux variation over the whole extended region cannot be completely excluded. If the flux variation were dominated by a smaller region in the source area, the picture could be more reasonable.
In such a scenario, however, identifying MGRO J2019+37 as a PWN could be a dilemma because it otherwise should have a steady flux.
\section{Cosmic Ray Physics}
Several interesting results have been obtained by ARGO-YBJ in the CR physics, as discussed in \cite{dettorre11}.
In the following sections the measurements of the anisotropy in the CR arrival direction distribution and of the light component (p+He) spectrum of CRs are described.

\subsection{Large Scale CR Anisotropy}
The observation of the CR large scale anisotropy by ARGO-YBJ is shown in Fig. \ref{fig:lsa} as a function of the primary energy up to about 25 TeV. 

%
\begin{myfigure}
\centerline{\resizebox{70mm}{!} {\includegraphics{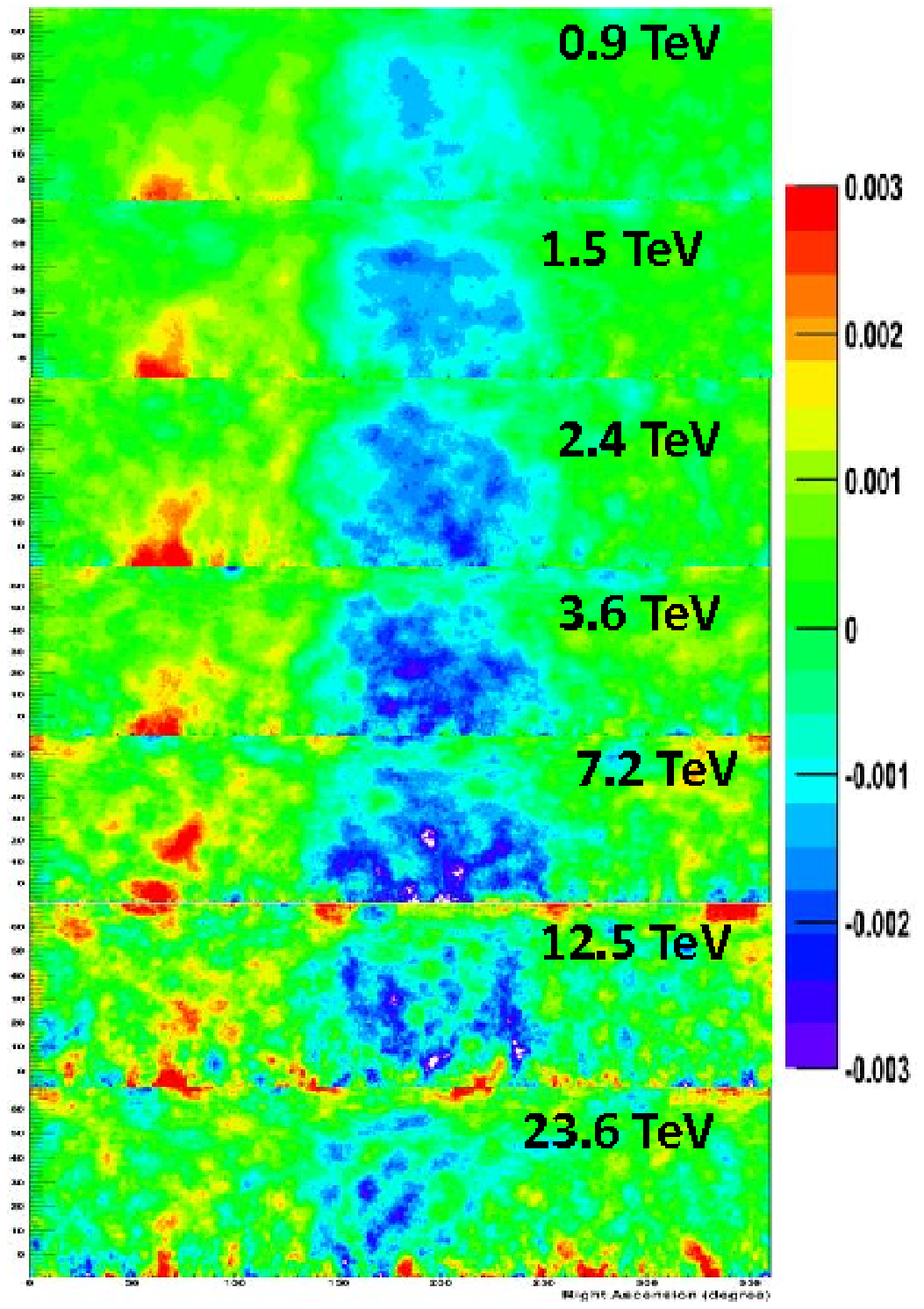}}}
\caption{Large scale CR anisotropy observed by ARGO-YBJ as a function of the energy. The color scale gives the relative CR intensity.}
\label{fig:lsa}
\end{myfigure}
%
%
\begin{myfigure}
\centerline{\resizebox{85mm}{!} {\includegraphics{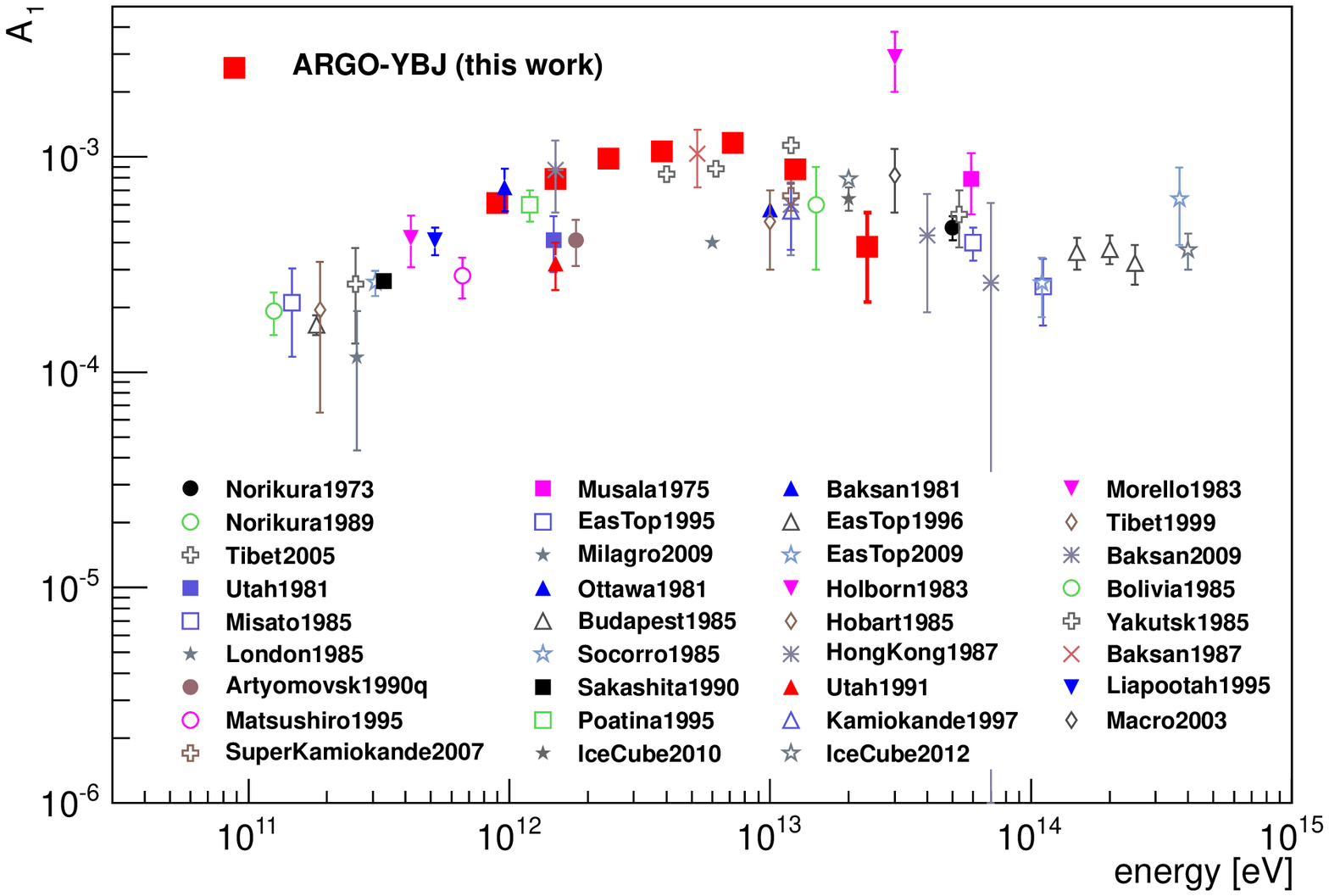}}}
\caption{Amplitude of the first harmonic as a function of the energy, compared with other measurements.}
\label{fig:lsa-ampl}
\end{myfigure}
%
%
\begin{myfigure}
\centerline{\resizebox{85mm}{!} {\includegraphics{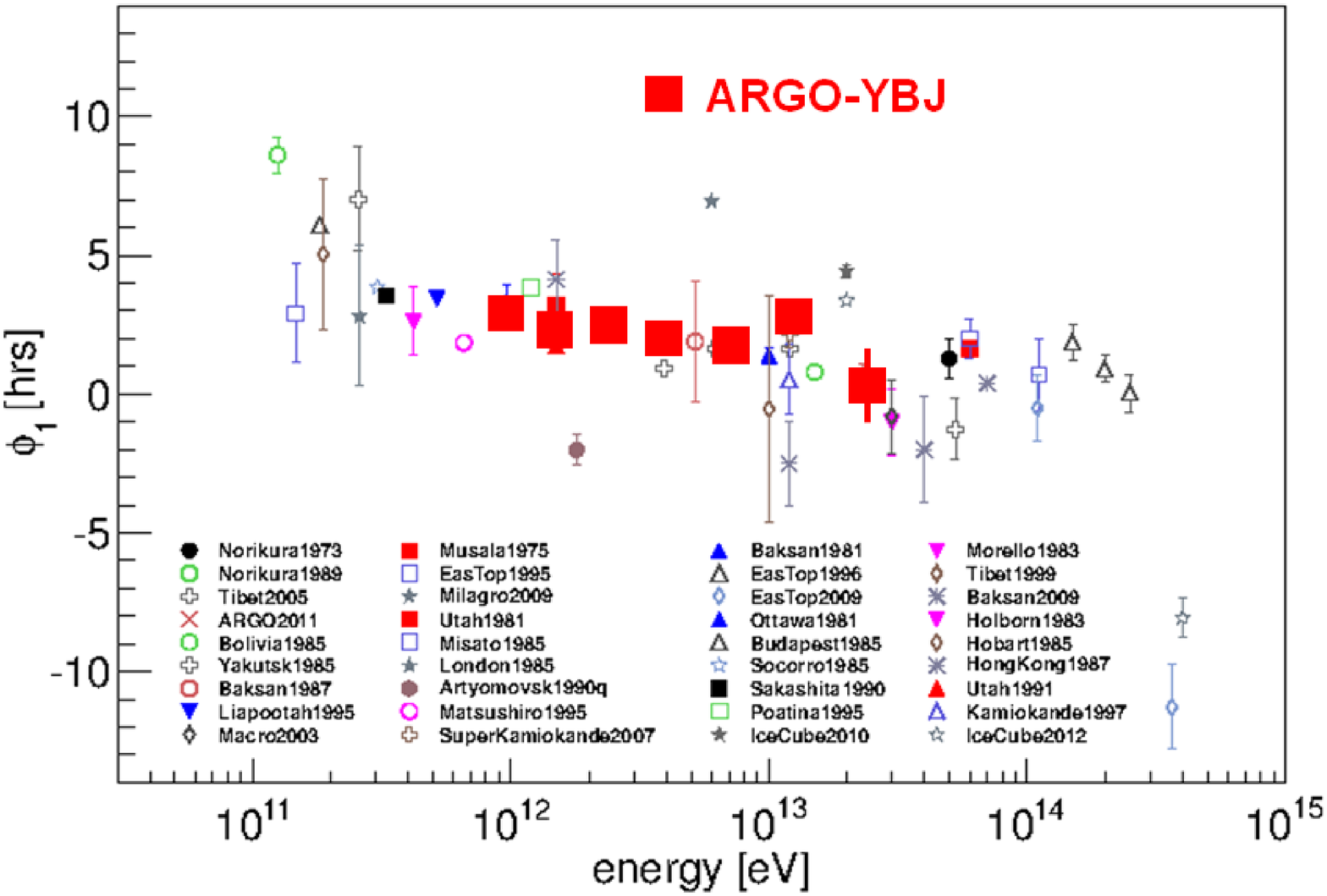}}}
\caption{Phase of the first harmonic as a function of the energy, compared with other measurements.}
\label{fig:lsa-phase}
\end{myfigure}
%
The so-called \textit{`tail-in'} and \textit{`loss-cone'} regions, correlated to an enhancement and a deficit of CRs, respectively, are clearly visible with a statistical significance greater than 20 s.d..
The tail-in broad structure appears to break up into smaller spots with increasing energy.
In order to quantify the scale of the anisotropy we studied the 1-D R.A. projections integrating the sky maps inside a declination band given by the field of view of the detector. Therefore, we fitted the R.A. profiles with the first two harmonics. The resulting amplitude and phase of the first harmonic are plotted in Fig. \ref{fig:lsa-ampl} and Fig. \ref{fig:lsa-phase} where are compared to other measurements as a function of the energy. The ARGO-YBJ results are in agreement those of other experiments, suggesting a decrease of the anisotropy first harmonic amplitude at energies above 10 TeV.
\subsection{Medium Scale Anisotropy}
Fig. \ref{fig:msa} shows the ARGO-YBJ sky map in equatorial coordinates containing about 2$\cdot$10$^{11}$ events reconstructed with a zenith angle $\leq$50$^{\circ}$.

%
\begin{myfigure}
\centerline{\resizebox{85mm}{!} {\includegraphics{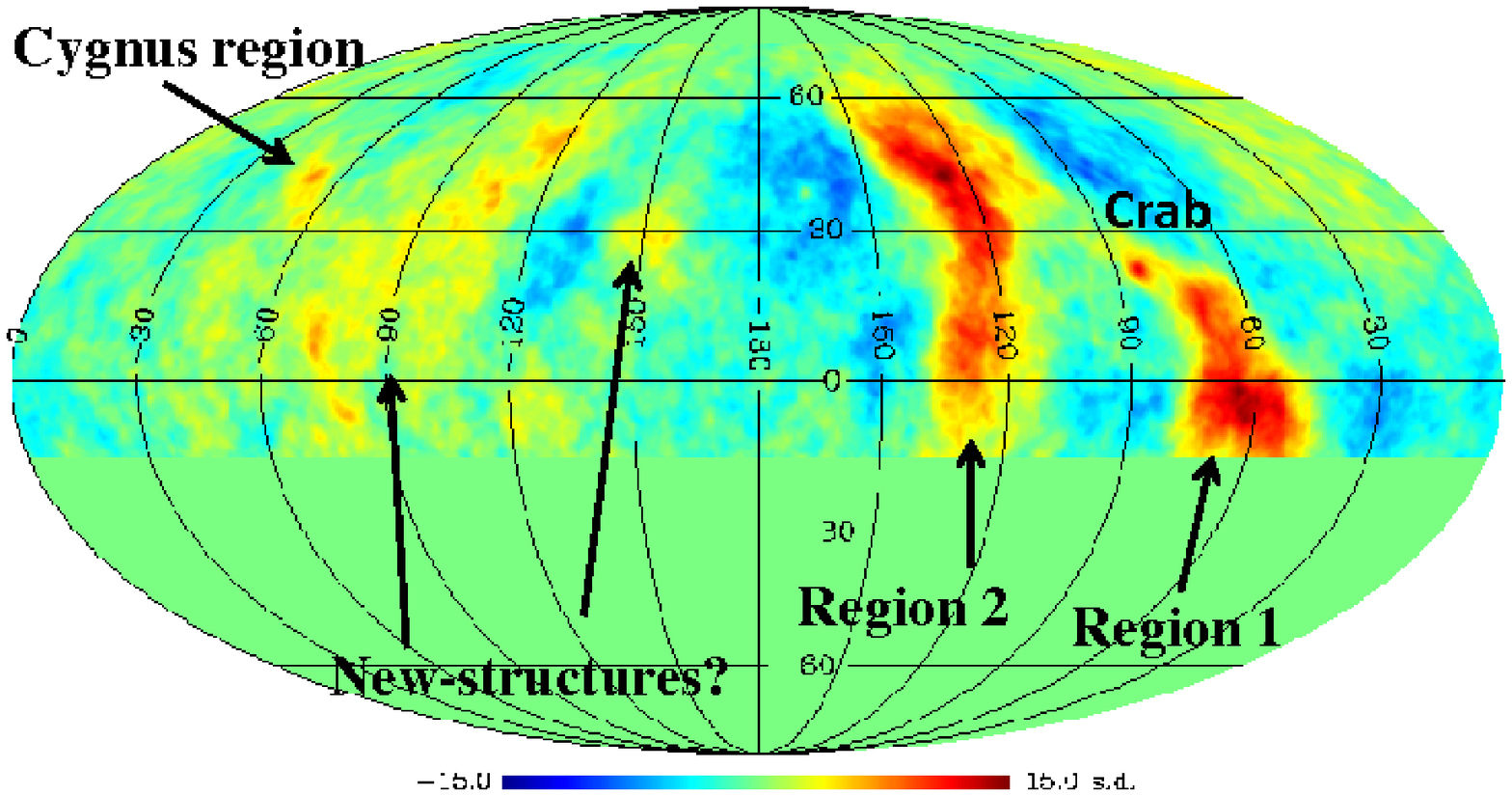}}}
\caption{Medium scale CR anisotropy observed by ARGO-YBJ. The color scale gives the statistical significance of the observation in standard deviations. }
\label{fig:msa}
\end{myfigure}
%
The zenith cut selects the declination region $\delta\sim$ -20$^{\circ}\div$ 80$^{\circ}$.
According to simulations, the median energy of the isotropic CR proton flux is E$_p^{50}\approx$1.8 TeV (mode energy $\approx$0.7 TeV).

The most evident features are observed by ARGO-YBJ around the positions $\alpha\sim$ 120$^{\circ}$, $\delta\sim$ 40$^{\circ}$ and $\alpha\sim$ 60$^{\circ}$, $\delta\sim$ -5$^{\circ}$, positionally coincident with the excesses detected by Milagro \cite{milagro08}. These regions, named ``1'' and ``2'', are observed with a statistical significance of about 14 s.d..  The deficit regions parallel to the excesses are due to a known effect of the analysis, that uses also the excess events to evaluate the background, overestimating this latter.

The left side of the sky map seems to be full of few-degree excesses not compatible with random fluctuations (the statistical significance is more than 6 s.d. post-trial). 
The observation of these structures is reported here for the first time and together with that of regions 1 and 2 it may open the way to an interesting study of the TeV CR sky.

To figure out the energy spectrum of the excesses, data have been divided into five independent shower multiplicity sets. The number of events collected within each region is computed for the event map (Ev) as well as for the background one (Bg). The relative excess (Ev-Bg)/Bg is computed for each multiplicity interval. 

%
\begin{myfigure}
\centerline{\resizebox{70mm}{!} {\includegraphics{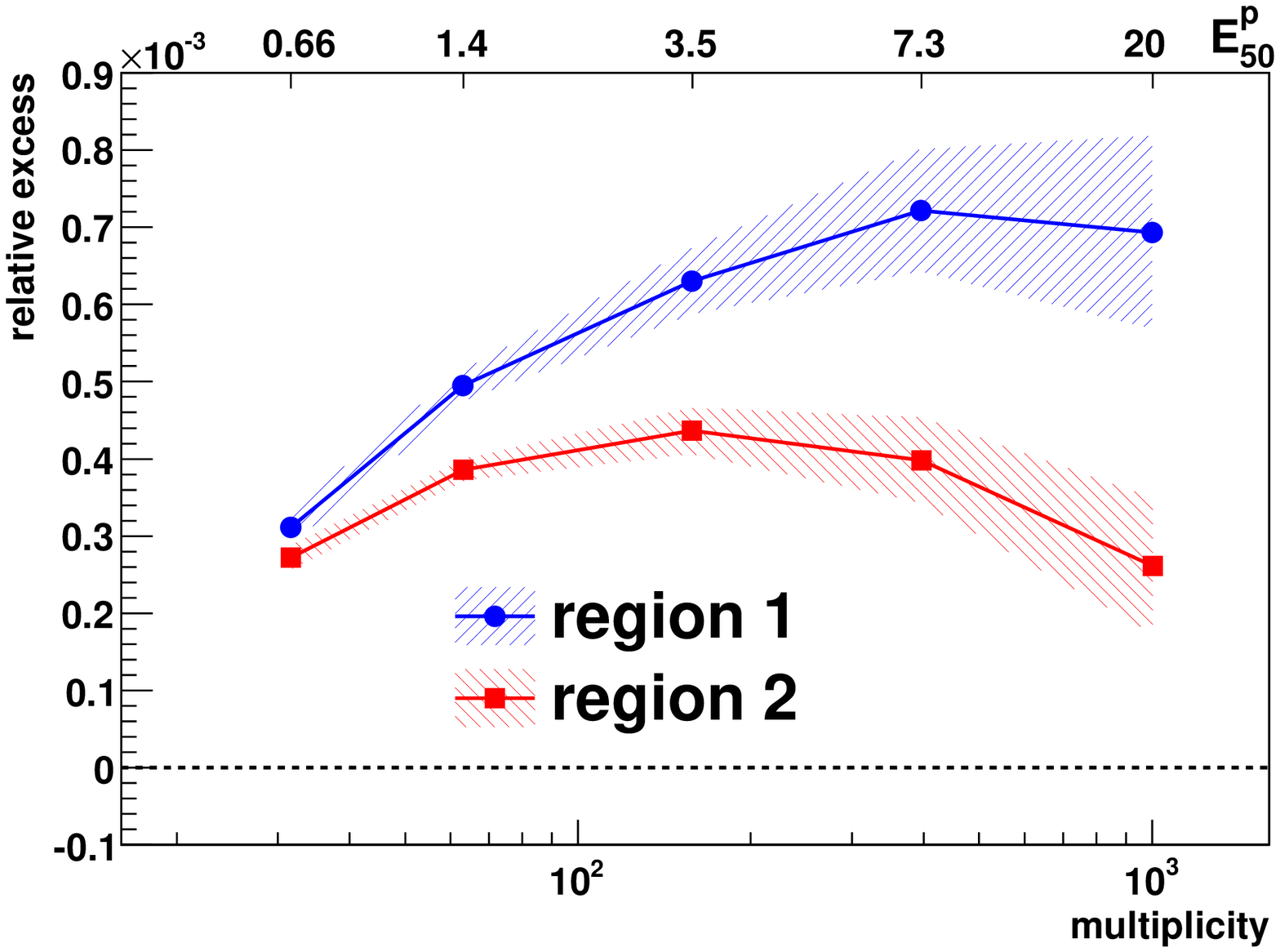}}}
\caption{Size spectrum of the regions 1 and 2. The vertical axis represents the relative excess (Ev-Bg)/Bg. The upper scale shows the corresponding proton median energy. The shadowed areas represent the 1$\sigma$ error band.}
\label{fig:msa-ensp}
\end{myfigure}
%
The result is shown in Fig. \ref{fig:msa-ensp}. Region 1 seems to have a spectrum harder than isotropic CRs and a cutoff around 600 shower particles (proton median energy E$^{50}_p$ = 8 TeV). On the other hand, the excess hosted in region 2 is less intense and seems to have a spectrum more similar to that of isotropic CRs.
We note that, in order to filter the global anisotropy, we used a method similar to that used by Milagro and Icecube. 
Further studies using different approaches are under way.
\subsection{Light component (p+He) spectrum of CRs}
%
\begin{myfigure}
\centerline{\resizebox{70mm}{!} {\includegraphics{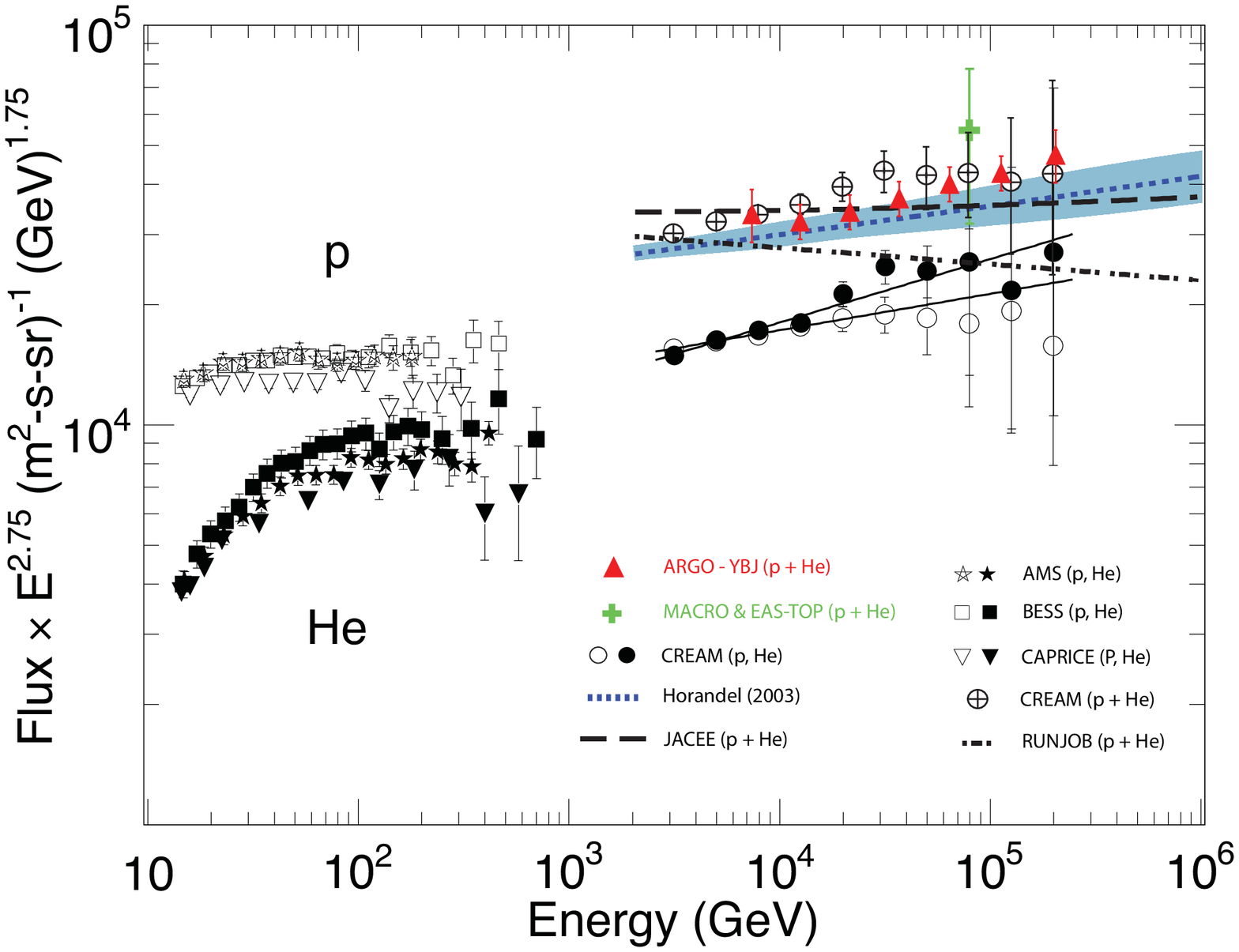}}}
\caption{Light component (p+He) spectrum of primary CRs measured by ARGO-YBJ compared with other experimental results.}
\label{fig:light_spectrum}
\end{myfigure}
%
Requiring quasi-vertical showers ($\theta<$30$^{\circ}$) and applying a selection criterion based on the particle density, a sample of events mainly induced by protons and helium nuclei with shower core inside a fiducial area (with radius $\sim$28 m) has been selected. The contamination by heavier nuclei is found negligible. An unfolding technique based on the Bayesian approach has been applied to the strip multiplicity distribution in order to obtain the differential energy spectrum of the light component (p + He nuclei) in the energy range (5 - 200) TeV \cite{bartoli12e}. 
The spectrum measured by ARGO-YBJ is compared with other experimental results in Fig. \ref{fig:light_spectrum}.  
Systematic effects due to different hadronic models and to the selection criteria do not exceed 10\%.
The ARGO-YBJ data agree remarkably well with the values obtained by adding up the p and He fluxes measured by CREAM both concerning the total intensities and the spectral index \cite{cream11}. The value of the spectral index of the power-law fit to the ARGO-YBJ data is -2.61$\pm$0.04, which should be compared with $\gamma_p$ = -2.66$\pm$0.02 and $\gamma_{He}$ = -2.58$\pm$0.02 obtained by CREAM.
The present analysis does not allow the determination of the individual p and He contribution to the measured flux, but the ARGO-YBJ data clearly exclude the RUNJOB results \cite{runjob}. 
We emphasize that for the first time direct and ground-based measurements overlap for a wide energy range thus making possible the cross-calibration of the different experimental techniques.
\section{Conclusions}
The ARGO-YBJ detector exploiting the full coverage approach and the high segmentation of the readout is imaging the front of atmospheric showers with unprecedented resolution and detail. The digital and analog readout will allow a deep study of the CR phenomenology in the wide TeV - PeV energy range. The results obtained in the low energy range (below 100 TeV) predict an excellent capability to address a wide range of important issues in Astroparticle Physics.

\section{Dedication}

This article is dedicated to the memory of my father who passed away on September 11th, 2012.

\end{multicols}
\end{document}